\documentclass[aps,prb,superscriptaddress,twocolumn,amsmath,amssymb,titlepage]{revtex4-1}

\usepackage{bbold}
\usepackage{mathptmx}
\usepackage{psfrag,graphicx}
\usepackage{dcolumn}
\usepackage{amsmath,amssymb}
\usepackage{bm}
\usepackage{color}
\usepackage{latexsym}
\usepackage{epstopdf}
\usepackage{color}
\usepackage[english]{babel}
\usepackage{latexsym}
\usepackage{psfrag,graphicx}
\usepackage{subfig}
\usepackage{amsmath}
\usepackage{amssymb}
\usepackage{amsfonts}
\usepackage{bm}
\usepackage{natbib}
\usepackage{epstopdf}
\usepackage{appendix}
\usepackage{rotating}

\definecolor{mygrey}{gray}{0.35}
\definecolor{myblue}{rgb}{0.2,0.2,0.8}
\definecolor{myzard}{cmyk}{0,0,0.05,0}
\definecolor{mywhite}{rgb}{1,1,1}
\definecolor{mywhite}{rgb}{1,1,1}
\definecolor{myred}{rgb}{1,0.,0.3}

\usepackage[colorlinks=true,citecolor=myblue,linkcolor=myred]{hyperref}

\def\ba{\begin{align}}
\def\enda{\end{align}}
\def\bi{\begin{itemize}}
\def\ei{\end{itemize}}

\def\be{\begin{equation}}
\def\ee{\end{equation}}
\def\bea{\begin{eqnarray}}
\def\eea{\end{eqnarray}}
\def\bse{\begin{subequations}}
\def\ese{\end{subequations}}

\newcommand{\ket}[1]{|{#1}\rangle}                       
\newcommand{\average}[1]{\langle {#1} \rangle}           

\newcommand{\Ignore}[1]{ }

\begin{document}

\title{Coupling-assisted Landau-Majorana-St\"uckelberg-Zener transition in two-interacting-qubit systems}

\author{R. Grimaudo}
 \affiliation{ Dipartimento di Fisica e Chimica dell'Universit\`a di Palermo, Via Archirafi, 36, I-90123 Palermo, Italy}
\affiliation{ INFN, Sezione Catania, \textit{I-95123} Catania, Italy}

\author{ N. V. Vitanov }
\affiliation{Department of Physics, St. Kliment Ohridski University of Sofia, 5 James Bourchier Boulevard, 1164 Sofia, Bulgaria}

\author{A. Messina}
\affiliation{ INFN, Sezione Catania, \textit{I-95123} Catania, Italy}
\affiliation{ Dipartimento di Matematica ed Informatica dell'Universit\`a di Palermo, Via Archirafi, 34, I-90123 Palermo, Italy}

\begin{abstract}
We analyse a system of two interacting spin-qubits subjected to a Landau-Majorana-St\"uckelberg-Zener (LMSZ) ramp.
We prove that LMSZ transitions of the two spin-qubits are possible without an external transverse static field since its role is played by the coupling between the spin-qubits.
We show how such a physical effect could be exploited to estimate the strength of the interaction between the two spin-qubits and to generate entangled states of the system by appropriately setting the slope of the ramp.
Moreover, the study of effects of the coupling parameters on the time-behaviour of the entanglement is reported.
Finally, our symmetry-based approach allows us to discuss also effects stemming from the presence of a classical noise or non-Hermitian dephasing terms. 
\end{abstract}

\date{\today}

\pacs{ 
PACS 
}

\maketitle

\section{Introduction}

The Landau-Majorana-St\"uckelberg-Zener (LMSZ) scenario \cite{LMSZ} and the Rabi one \cite{Rabi1937} represent two milestones among exactly solvable time-dependent semi-classical models for two-level systems.
A common fundamental property of these two models is the possibility of realizing a full population inversion in a two-state quantum system.
In the former case through an adiabatic passage via a level crossing, in the second case thanks to the application of a resonant $\pi$-pulse.

It is important to underline that the LMSZ scenario, differently from the Rabi case, is an ideal model.
The word ``ideal'' refers to the fact that it consists in a process characterized by an infinite time duration resulting, then, practically unrealisable.
This fact leads, indeed, to not physical properties such as, for example, the fact that the energies of the adiabatic states diverge at initial ($-\infty$) and final instant ($+\infty$).
As a consequence, both mathematical and physical problems arise when amplitudes and not only probabilities are necessary, e.g. when initial states present coherences  \cite{Torosov,Vasilev}.
In such cases one can alternatively use either the exact solutions of the finite LMSZ scenario \cite{Vit-Garr} or the Allen-Eberly-Hioe model \cite{AE}, the Demkov-Kunike model \cite{DK} or other models \cite{Vasilev1,Letho}, where no divergency problems arise and the transition probability is rather simple.

However, despite this circumstance, it is a matter of fact that the LMSZ grasps peculiar dynamical aspects of a lot of physical systems \cite{Shevchenko1}.
This relevant aspect has increased the popularity of the LMSZ model and several efforts have been done towards its generalization to the case of $N$-level quantum systems \cite{Vasilev,SSIvanov,Sinitsyn} and total crossing of bare energies \cite{Militello1}.
Moreover, its experimental feasibility gave it a basic role in the area of quantum technology thanks also to the several sophisticated techniques developed for a precise local manipulation of the state and the dynamics of a single qubit in a chain \cite{Wieser,Sivkov,Thiele,Reilly,Huang,Randall}.

In such an applicative scenario, as we know, several sources of incoherences can be present \cite{Petta,Anderlini,Foletti,Das Sarma Nat}: incoherent (mixed) states, relaxation processes (e.g., spontaneous emission) or interaction with a surrounding environment (e.g., nuclear spin bath).
They generate incoherent excitation leading to departure from a perfect (ideal) population transfer.
Therefore, more realistic descriptions of quantum systems subjected to LMSZ scenario comprising such effects have been proposed \cite{Akulin,Vitanov,Ivanov,Pok-Sin2003,Militello2,Militello3}.

In this respect, the most relevant influence in the dynamics of a spin-qubit primarily stems from the coupling with its nearest neighbours.
Recently the attention has been focused on double interacting spin-qubit systems subjected to LMSZ scenario \cite{VitPRL2001,Ribeiro,Ribeiro1,Rancic,Larson,Shevchenko}.
These papers investigate the coupling effects in the two-spin system dynamics in view of possible experimental techniques and protocols.
Moreover, such systems, under specific conditions, behave effectively as a two-level system with relevant applicability in quantum information and computation sciences \cite{Mhel}.
In the references cited before, indeed, generation of entangled states \cite{VitPRL2001} or the singlet-triplet transition \cite{Reilly,Ribeiro,Ribeiro1} in the two-qubit system under the LMSZ scenario have been studied.

With the same objective in mind, that is to characterize physical effects stemming from the coupling between two spin-qubits subjected to a LMSZ scenario, in this paper we study a two-spin-1/2 system described by a $C_2$-symmetry Hamiltonian model.
We consider coupling terms compatible with the symmetry of the Hamiltonian, namely isotropic and anisotropic exchange interaction.
The two spin-1/2's are moreover subjected only to a LMSZ ramp with no transverse static field.
We show that LMSZ transitions for the two spin-qubits are still possible thanks to the presence of the coupling, playing the role of an effective transverse field.
Such an effect, we call \textit{coupling-assisted LMSZ transition}, deserves particular attention for two reasons.
Firstly, it can be exploited to estimate the presence and the relative weight of different coupling parameters determining the symmetry of the Hamiltonian and then the dynamics of the two spins.
Secondly, through such an estimation, it is possible to set the slope of the field ramp in such a way to generate asymptotic entangled states of the two qubits.

The paper is organized as follows.
In Sec. \ref{Sec II} we introduce the model and its symmetry properties on which the dynamical reduction is based.
In Sec. \ref{Sec III} the application of the LMSZ scenario on both the subdynamics (that is the two-qubit dynamics restricted to the invariant subspaces) is performed.
Moreover, physical effects stemming from the (an)isotropy of the exchange interaction are brought to light.
In the subsequent Sec. \ref{Sec IV}, we emphasize the possibility of estimating the values of the coupling parameters. The generation of asymptotic entangled states of the two spins through coupling-based LMSZ transitions is reported instead in Sec. \ref{Sec IV-V}.
Some effects of a possible interaction with a surrounding environment, providing for either a classical noisy field component or non-Hermitian terms in the Hamiltonian model, are taken into account in Sec. \ref{Sec V}.
Finally, some conclusive comments and further remarks can be found in the last Sec. \ref{Sec C}.

\section{The Model}\label{Sec II}

Let us consider the following model, describing two interacting spin-qubits:
\begin{equation} \label{Hamiltonian}
\begin{aligned}
{H} =
\hbar\omega_{1}(t)\hat{\sigma}_{1}^{z}+\hbar\omega_{2}(t)\hat{\sigma}_{2}^{z}+\gamma_{x}\hat{\sigma}_{1}^{x}\hat{\sigma}_{2}^{x}+\gamma_{y}\hat{\sigma}_{1}^{y}\hat{\sigma}_{2}^{y}+\gamma_{z}\hat{\sigma}_{1}^{z}\hat{\sigma}_{2}^{z}
\end{aligned}
\end{equation}
where $\hat{\sigma}_{i}^{x}$, $\hat{\sigma}_{i}^{y}$ and $\hat{\sigma}_{i}^{z}$ ($i=1,2$) are the Pauli matrices and all the parameters may be thought as time-dependent.
The matrices are represented in the following ordered two-spin basis $\{ \ket{++},\ket{+-},\ket{-+},\ket{--} \}$ $(\hat{\sigma}^z\ket{\pm}=\pm\ket{\pm})$.

The $C_2$-symmetry with respect to the $z$-direction, possessed by the Hamiltonian, causes the existence of two dynamically invariant Hilbert subspaces related to the two eigenvalues of the constant of motion $\hat{\sigma}_1^z\hat{\sigma}_2^z$ \cite{GMN}.
Basing on such a symmetry, the time evolution operator, solution of the Schr\"odinger equation $i\hbar\dot{U}=HU$, may be formally put in the following form \cite{GMN}
\begin{equation}\label{Total time ev op}
U =
\begin{pmatrix}
a_+(t) & 0 & 0 & b_+(t) \\
0 & a_-(t) & b_-(t) & 0 \\
0 & -b_-^*(t) & a_-^*(t) & 0 \\
-b_+^*(t) & 0 & 0 & a_+^*(t) \\
\end{pmatrix}.
\end{equation}
The condition $U(0)=\mathbb{1}$ is satisfied by putting $a_\pm(0)=1$ and $b_\pm(0)=0$.
It is worth noticing that $a_\pm(t)$ and $b_\pm(t)$ are the time-dependent parameters of the two evolution operators
\begin{equation}\label{Time Ev Ops 2x2}
U_{\pm} = e^{\mp i \gamma_{z} t / \hbar}
\begin{pmatrix}
a_\pm(t) & b_\pm(t) \\
-b_\pm^*(t) & a_\pm^*(t)
\end{pmatrix},
\end{equation}
solutions of two independent dynamical problems of fictitious single spin-1/2, namely $i\hbar \dot{U}_\pm = H_\pm U_\pm$, $U_\pm(0)=\mathbb{1}_\pm$, with
\begin{equation}
\begin{aligned}
H_{\pm}=&
\begin{pmatrix}
\hbar\Omega_{\pm}(t) & \gamma_{\pm} \\
\gamma_{\pm} & -\hbar\Omega_{\pm}(t)
\end{pmatrix}
\pm \gamma_{z} \mathbb{1}_\pm \\
=&\hbar\Omega_\pm(t)\hat{\sigma}^z+\gamma_\pm\hat{\sigma}^x\pm\gamma_z\mathbb{1}_\pm,
\end{aligned}
\end{equation}
where
\begin{equation} \label{Omega+- and gamma+-}
\begin{aligned}
\Omega_{\pm}(t) = [\omega_{1}(t) \pm \omega_{2}(t)], \qquad
\gamma_{\pm} = (\gamma_{x} \mp \gamma_{y}),
\end{aligned}
\end{equation}
and $\mathbb{1}_\pm$ represent the identity operators within the two-dimensional subspaces.
Thus, it means that the solution of the dynamical problem of the two interacting spin-1/2's is traced back to the solution of two independent problems, each one of single (fictitious) spin-1/2 \cite{GMN}. 

The explicit expressions of $a_\pm(t)$ and $b_\pm(t)$ depend on the specific time-dependences of $\omega_1(t)$ and $\omega_2(t)$.
It is well known that it is not possible to find the analytical solution of the Schr\"odinger equation for a spin-1/2 subjected to a generic time-dependent field.
Therefore, specific exactly solvable time-dependent scenarios for a single spin-1/2 might be of great help to investigate the dynamics of the two interacting spin system under scrutiny \cite{GMN}.

\section{Coupling-based LMSZ Transition}\label{Sec III}

In this section we investigate the case in which a LMSZ ramp is applied on either just one or both the spins.
Our following theoretical analysis is based on the possibility of experimentally addressing at will the spin systems exploiting, for example, the Scanning Tunneling Microscopy (STM).
It appears hence appropriate to furnish a sketch of such a technique.

STM proved to be an excellent experimental technique in controlling the dynamics of spin-qudit systems for two main reasons: 1) the possibility of building atom by atom atomic-scale structures \cite{Khajetoorians}, such as spin chains and nano-magnets \cite{Yan}; 2) the possibility of controlling the whole system by addressing a single element (qudit) while it interacts with the others \cite{Yan,Bryant,Tao}, succeeding in realizing, for example, logic operations \cite{Khajetoorians}.
The manipulation of a single qudit dynamics is performed through the exchange interaction between the atom on the tip of the scanning tunneling microscope and the target atom in the chain.
It is possible to show that such an interaction is equivalent to a magnetic field applied on the atom we want to manipulate \cite{Yan,Wieser}.
In this way, it is easy to guess that a time-dependent distance between the tip and the target atom generates a time-dependent exchange coupling, giving rise, in turn, to a time-dependent effective magnetic field on the atom of the chain, as analysed in Ref. \cite{Wieser}.
Basing on such an observation, in Ref. \cite{Sivkov} the authors study the spin dynamics and entanglement generation in a spin chain of Co atoms on a surface of $\text{Cu}_3$N/Cu(110).
Precisely, they consider a LMSZ ramp along the $z$ direction produced in a time window of $20 ps$ and a short Gaussian pulse in the $x$ direction (half-width: $10 ps$).

\subsection{Collective LMSZ Dynamics}

At the light of the STM experimental scenario, we take into account firstly the case of a LMSZ ramp applied on the first spin such that
\begin{equation}
\hbar\omega_1(t)={\alpha} t/2, \quad \hbar\omega_2(t)=0, \qquad t\in (-\infty,\infty),
\end{equation}
where $\alpha$ is related to the velocity of variation of the field, $\dot{B}_z\propto\alpha$, and it is considered a positive real number without loss of generality.
Let us consider, moreover, the two spins initialized in the state $\ket{--}$.
In this instance, the subdynamics governed by $H_+$ is characterized by a LMSZ scenario where the longitudinal ($z$) magnetic field produces the standard LMSZ ramp $\hbar\Omega_+(t)=\hbar\omega_1(t)=\alpha t/2$ and the transverse effective magnetic field along the $x$-direction is given by $\gamma_-$.
It is well-known that the dynamical problem for such a time-dependent scenario can be analytically solved.
The transition probability of finding the two-spin system in the state $\ket{++}$ coincides with the probability of finding the fictitious spin-1/2 subjected to $H_+$ in its state $\ket{+}$ starting from $\ket{-}$ and reads \cite{LMSZ}
\begin{equation}\label{P+ simple}
P_+=|\average{++|U_+(\infty)|--}|^2=1-\exp\{ -2\pi\gamma_+^2/\hbar\alpha \}.
\end{equation}

If we now, instead, consider the two spins initially prepared in $\ket{-+}$, the probability for each spin-1/2 of undergoing a LMSZ transition, that is the probability of finding the two-spin system in the state $\ket{+-}$, results
\begin{equation}\label{P- simple}
P_-=|\average{+-|U_-(\infty)|-+}|^2=1-\exp\{ -2\pi\gamma_-^2/\hbar\alpha \}.
\end{equation}
This time the transition probability is governed by the fictitious magnetic field given by $\gamma_-$.
The effective longitudinal magnetic field, instead, is the same, namely $\hbar\Omega_-(t)=\hbar\omega_1(t)=\alpha t/2$.
We see that in both cases, though a constant transverse magnetic field is absent, the LMSZ transition of both the spins is possible thanks to the presence of the coupling between them.
It is important to stress that, for the cases considered before, if $\gamma_x = \gamma_y$ (as it often happens experimentally) we cannot have transition in the first case, that is in the subdynamics involving $\ket{++}$ and $\ket{--}$.
In this instance, indeed, $P_+$ happens to be 0 at any time.

\subsection{Isotropy Effects: Local LMSZ Transition by nonlocal Control and State Transfer}\label{Part In Cond}

The symmetry-based dynamical decomposition and the application of the STM LMSZ scenario in each subdynamics allow us to bring to light peculiar evolutions of physical interest.
For example, if we consider $\gamma_x \neq \gamma_y$ and the following initial condition
\begin{equation}\label{Particular In St 1}
\ket{+} \otimes {\ket{+}+\ket{-} \over \sqrt{2}},
\end{equation}
the two states $\ket{++}$ and $\ket{--}$ evolve independently and applying the LMSZ ramp we have the probability $P=P_+P_-$ to find asimptotically the two-spin system in the state
\begin{equation}\label{Asymptotic In St 1 anisotropy}
\ket{-} \otimes {\ket{+}+\ket{-} \over \sqrt{2}}.
\end{equation}
We see, that such a dynamics leaves unaffected the second spin, while it produces a LMSZ transition only on the first spin.
It is also relevant the dynamical evolution of the symmetric initial condition
\begin{equation}\label{Particular In St 2}
{\ket{+}+\ket{-} \over \sqrt{2}} \otimes \ket{+}.
\end{equation}
This time, we get the same probability $P=P_+P_-$ of finding asymptotically the two-spin system in
\begin{equation}\label{Asymptotic In St 2 anisotropy}
{\ket{+}+\ket{-} \over \sqrt{2}} \otimes \ket{-}.
\end{equation}
This case results less intuitive even if we are reproducing the same dynamics but with interchanged roles of the two spins.
In this instance, in fact, we generate a LMSZ transition only on the second spin by locally applying the field on the first one.
This shows that the coupling between the two spins plays a key role to achieve a non-local control of the second spin by locally manipulating the first ancilla qubit.

If we consider, instead, $\gamma_x=\gamma_y=\gamma/2$ we know that the transition $\ket{--} \leftrightarrow \ket{++}$ is suppressed.
This means that if we consider as initial conditions the states in Eqs. \eqref{Particular In St 1} and \eqref{Particular In St 2}, we get asymptotically, this time, the states
\begin{subequations}
\begin{align}
&{\ket{+}+\ket{-} \over \sqrt{2}} \otimes \ket{+}, \label{Asymptotic In St 1 isotropy} \\
&\ket{+} \otimes {\ket{+}+\ket{-} \over \sqrt{2}}, \label{Asymptotic In St 2 isotropy}
\end{align}
\end{subequations}
respectively, with probability $P=1-\exp\{ -2\pi\gamma^2/\hbar\alpha \}$.
We see that the isotropy properties of the exchange interaction consistently change the dynamics of the system.
When the exchange interaction is isotropic, indeed, the asymptotic states reached by the initial conditions \eqref{Particular In St 1} and \eqref{Particular In St 2} radically change.
In these cases, the resulting physical effect is  a state transfer or a state exchange between the two spin-qubits.
Therefore, the different state transitions from the state \eqref{Particular In St 1} [\eqref{Particular In St 2}] to the state \eqref{Asymptotic In St 1 anisotropy} or \eqref{Asymptotic In St 1 isotropy} [\eqref{Asymptotic In St 2 anisotropy} or \eqref{Asymptotic In St 2 isotropy}] can reveal the level of isotropy of the exchange interaction.

\section{Coupling Parameter Estimation}\label{Sec IV}

It is interesting noticing that the coupling-based LMSZ transition could be used to estimate the coupling parameters.
By measuring $P_+$ and $P_-$ (Eqs. \eqref{P+ simple} and \eqref{P- simple}, respectively) in a physical scenario describable by the Hamiltonian model \eqref{Hamiltonian}, we get an estimation of $\gamma_+$ and $\gamma_-$ and then of the two coupling parameters $\gamma_x$ and $\gamma_y$.
Supposing to know $P_+$ and $P_-$, we have indeed
\begin{equation}
\begin{aligned}
\gamma_x &= {1 \over 2}\sqrt{{\hbar\alpha \over 2\pi}}\left[ \sqrt{\log\left( {1 \over 1-P_-} \right)} + \sqrt{\log\left( {1 \over 1-P_+} \right)} \right], \\
\gamma_y &= {1 \over 2}\sqrt{{\hbar\alpha \over 2\pi}}\left[ \sqrt{\log\left( {1 \over 1-P_-} \right)} - \sqrt{\log\left( {1 \over 1-P_+} \right)} \right].
\end{aligned}
\end{equation}
We wish to emphasize that we may estimate the coupling parameters also through the Rabi oscillations occurring in the two subspaces.
Applying, indeed, a constant field $\omega_1$ on the first spin, the two probabilities $P_+$ and $P_-$ become
\begin{equation}
\begin{aligned}
&P_+= {\gamma_+^2 \over \hbar^2\omega_1^2+\gamma_+^2} \sin^2\left(\sqrt{\omega_1^2+\gamma_+^2/\hbar^2}~t\right), \\
&P_-= {\gamma_-^2 \over \hbar^2\omega_1^2+\gamma_-^2} \sin^2\left(\sqrt{\omega_1^2+\gamma_-^2/\hbar^2}~t\right).
\end{aligned}
\end{equation}
So, by measuring the frequency and the amplitude of the oscillations in the two cases we may get information about the the relative weights of the coupling parameters.

\section{Entanglement}\label{Sec IV-V}

\begin{figure}[htp]
\begin{center}
\subfloat[][]{\includegraphics[width=0.22\textwidth]{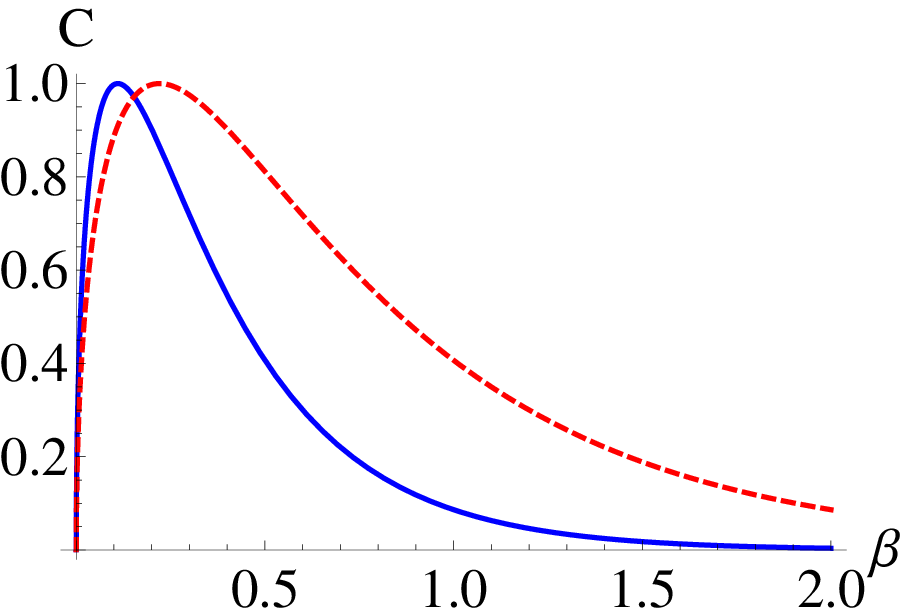}\label{fig:C1}}
\qquad
\subfloat[][]{\includegraphics[width=0.22\textwidth]{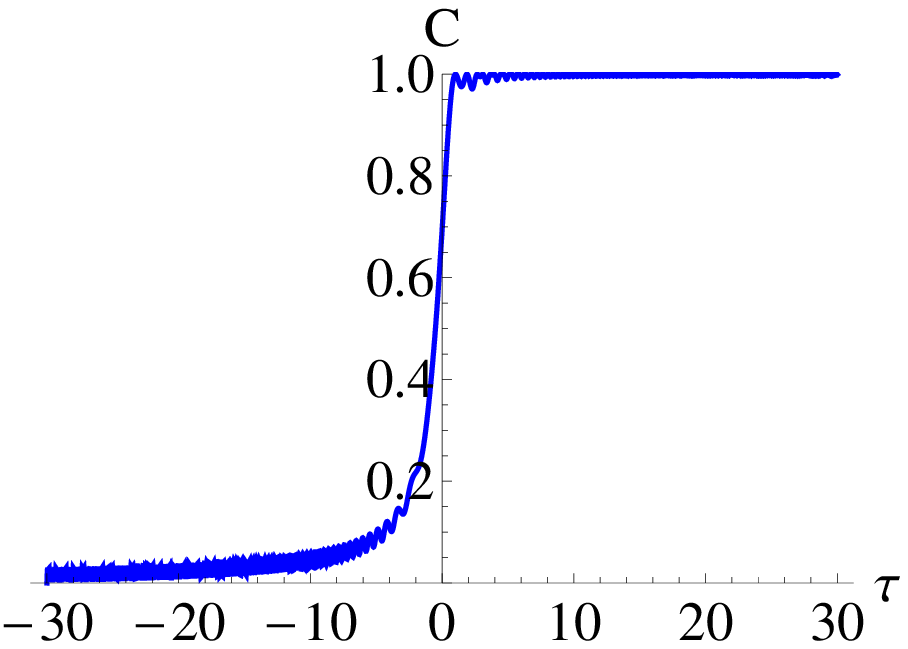}\label{fig:Cto1}}
\captionsetup{justification=raggedright,format=plain,skip=4pt}%
\caption{(Color online) a) The two curves of the concurrence in Eq. \eqref{C1a} (full blue line) and Eq. \eqref{C1b} (red dashed line) for $\beta_-/2=\beta_+=\beta$; b) Time behaviour of concurrence for the initial condition $\ket{--}$ and $\beta_+=0.1$ plotted against the dimensionless time $\tau=\sqrt{\alpha/\hbar}~t$.}
\end{center}
\end{figure}

A precise estimation of the coupling parameters is useful also to generate entangled states of the two spins.
By the knowledge of them, indeed, we may set the parameter $\alpha$ in order to get asymptotically $P_\pm=1/2$, generating so an entangled state.
If the two spins start from state $\ket{--}$ or $\ket{-+}$, being the dynamics unitary, they reach asymptotically the pure state $(\ket{++}+e^{i\phi}\ket{--})/\sqrt{2}$ in the first case and $(\ket{+-}+e^{i\phi}\ket{-+})/\sqrt{2}$ in the second case, which are maximally entangled states.
The asymptotic curves of the concurrence (the entanglemnt measure for two spin-1/2's introduced in Ref. \cite{Wootters}), in fact, when the two-spin system is initialized in $\ket{--}$ or $\ket{-+}$, read respectively
\begin{subequations}
\begin{align}
C&=2|c_{++}c_{--}|=2\sqrt{P_+(1-P_+)}=2\sqrt{(1-e^{-2\pi\beta_+})e^{-2\pi\beta_+}}\label{C1a} \\
C&=2|c_{+-}c_{-+}|=2\sqrt{P_-(1-P_-)}=2\sqrt{(1-e^{-2\pi\beta_-})e^{-2\pi\beta_-}}\label{C1b}
\end{align}
\end{subequations}
and they exhibit a maximum for $\beta_+=\beta_-=\log(2)/2\pi\approx 0.11$.
In the previous expressions we put $\beta_+=\gamma_+^2/\hbar\alpha$ and $\beta_-=\gamma_-^2/\hbar\alpha$, while $c_{++}$ and $c_{--}$ ($c_{+-}$ and $c_{-+}$) are the asymptotic amplitudes of the states $\ket{++}$ and $\ket{--}$ ($\ket{+-}$ and $\ket{-+}$), respectively.
Therefore, $\log(2)/2\pi$ is exactly the value the LMSZ parameters $\beta_+$ ad $\beta_-$ must have to realize the generation of the entangle states $(\ket{++}+e^{i\phi}\ket{--})/\sqrt{2}$ and $(\ket{+-}+e^{i\phi}\ket{-+})/\sqrt{2}$ when the two spins start from $\ket{--}$ or $\ket{-+}$, respectively.
Figure \ref{fig:C1} reports the two curves for $\beta_-/2=\beta_+=\beta$.

\begin{figure}[htp]\label{fig:Ct1}
\begin{center}
\subfloat[][]{\includegraphics[width=0.22\textwidth]{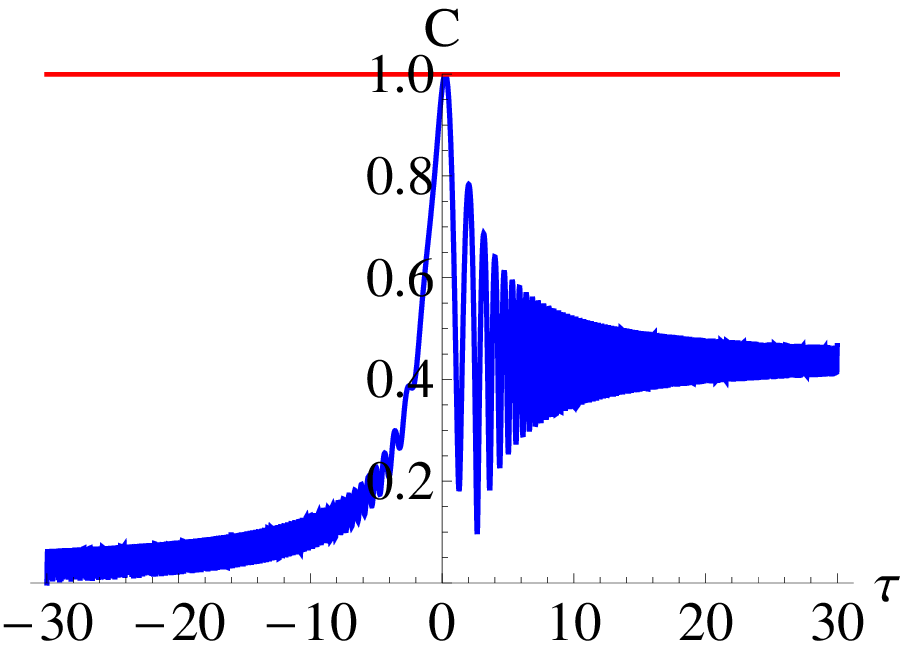}\label{fig:Cbeta05}}
\qquad
\subfloat[][]{\includegraphics[width=0.22\textwidth]{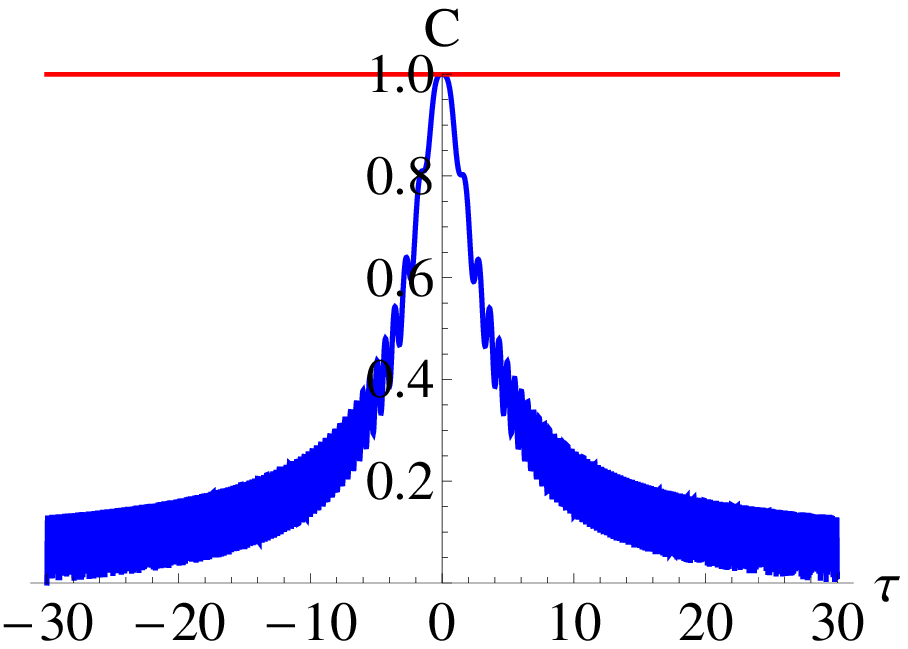}\label{fig:Cbeta2}}
\captionsetup{justification=raggedright,format=plain,skip=4pt}%
\caption{(Color online) Time behaviour of the concurrence against the dimensionless parameter $\tau=\sqrt{\alpha/\hbar}~t$ during a LMSZ process when the system starts from the state $\ket{--}$ for a) $\beta_+=1/2$ and b) $\beta_+=2$. The upper straight curve corresponds to $C(\tau)=1$.}
\end{center}
\end{figure}

We may verify this fact by investigating the behaviour of the concurrence in time.
To this end, the exact solutions of the two time-dependent parameters determining the two time evolution operators $U_+$ and $U_-$ in Eq. \eqref{Time Ev Ops 2x2}, related to each subdynamics, are necessary and they reads, namely \cite{Vit-Garr}
\begin{equation}\label{Exact a b}
\begin{aligned}
a_\pm=&{\Gamma_f(1-i\beta_\pm) \over \sqrt{2\pi}} \\
\times&[D_{i\beta_\pm}(\sqrt{2}e^{-i\pi/4}\tau) ~ D_{-1+i\beta_\pm}(\sqrt{2}e^{i3\pi/4}\tau_i) \\
&+D_{i\beta_\pm}(\sqrt{2}e^{i3\pi/4}\tau) ~ D_{-1+i\beta_\pm}(\sqrt{2}e^{-i\pi/4}\tau_i)],
\\\\
b_\pm=&{\Gamma_f(1-i\beta_\pm) \over \sqrt{2\pi\beta}} e^{i\pi/4} \\
\times&[-D_{i\beta_\pm}(\sqrt{2}e^{-i\pi/4}\tau) ~ D_{-1+i\beta_\pm}(\sqrt{2}e^{i3\pi/4}\tau_i) \\
&+D_{i\beta_\pm}(\sqrt{2}e^{i3\pi/4}\tau) ~ D_{-1+i\beta_\pm}(\sqrt{2}e^{-i\pi/4}\tau_i)].
\end{aligned}
\end{equation}
$\Gamma_f$ is the Gamma function, while $D_\nu(z)$ are the parabolic cylinder functions \cite{Abramowitz} and $\tau=\sqrt{\alpha/\hbar}~t$ is a time dimensionless parameter; $\tau_i$ identify the initial time instant.
If the system starts, e.g., from the state $\ket{--}$ the amplitudes result
\begin{equation}
c_{++}=b_+, \quad c_{--}=a_+^*, \quad c_{+-}=c_{-+}=0,
\end{equation}
and the related time-behaviour of the concurrence $C=2|b_+||a_+|$ for $\beta_+=0.1$ is reported in Fig. \ref{fig:Cto1}.
We see, as expected, that such a choice of the LMSZ parameter generate a maximally entangled state of the two spin-qubits.
It is important to point out that, on the basis of Eqs. \eqref{Exact a b}, the parameter $\beta_+$ determines not only the asymptotic value of the concurrence but also its time behaviour.
This fact is confirmed and can be appreciated by Figs. \ref{fig:Cbeta05} and Fig. \ref{fig:Cbeta2} reporting the concurrence against the dimensionless parameter $\tau$ for $\beta_+=1/2$ and $\beta_+=2$, respectively.
The physical meaning of the asymptotic vanishing of $C$ in Fig. \ref{fig:Cbeta2} is that for the specific value of $\beta_+$ the system evolves quite adiabatically towards the factorized states $\ket{--}$.
On the contrary, in Figs. \ref{fig:Cbeta05} the slope of the ramp induces a non adiabatic evolution towards a coherent not factorizable superposition of $\ket{++}$ and $\ket{--}$.

We would get analogous results by studying the LMSZ process when the two spin-qubits start from the state $\ket{-+}$.
In this case, only the states $\ket{-+}$ and $\ket{+-}$ would be involved and the LMSZ parameter determining the different concurrence regimes would be $\beta_-$.
For such initial conditions, then, the ratio $\beta_+/\beta_-$, imposing precise relationships between the coupling parameters $\gamma_x$ and $\gamma_y$, does not matter.

\begin{figure}[htp]\label{fig:Ct2}
\begin{center}
\subfloat[][]{\includegraphics[width=0.22\textwidth]{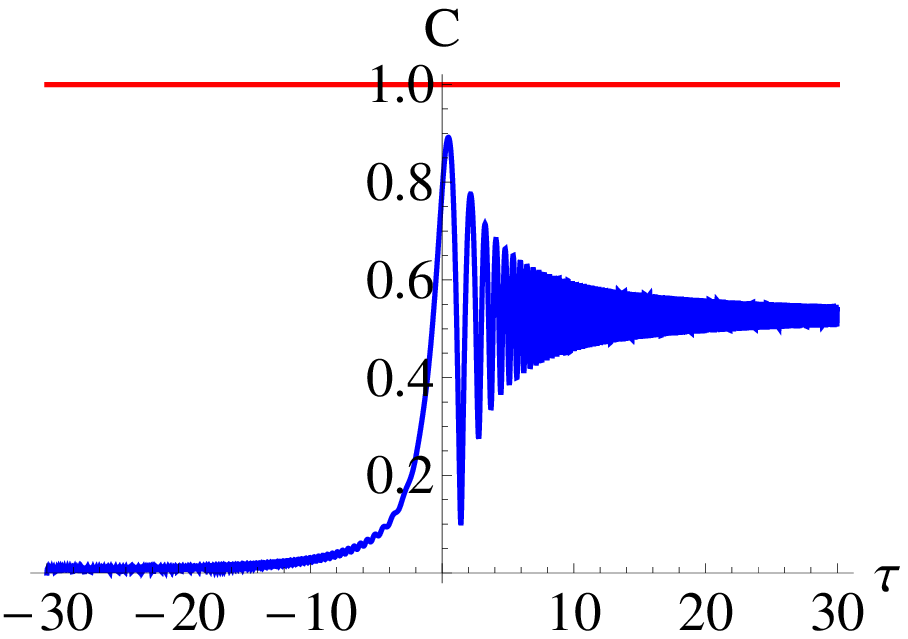}\label{fig:CB05beta05}}
\qquad
\subfloat[][]{\includegraphics[width=0.22\textwidth]{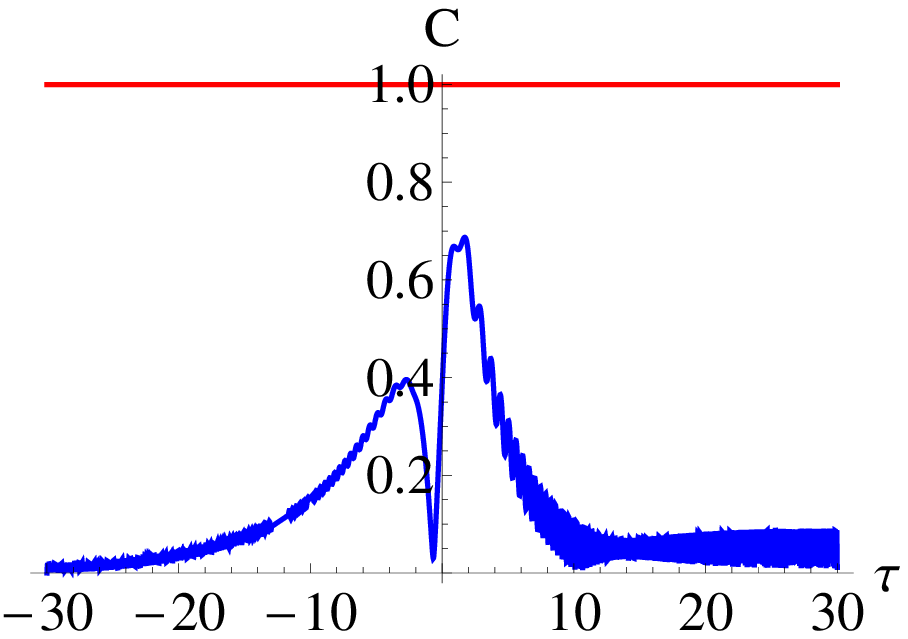}\label{fig:CB05beta2}}
\qquad
\subfloat[][]{\includegraphics[width=0.22\textwidth]{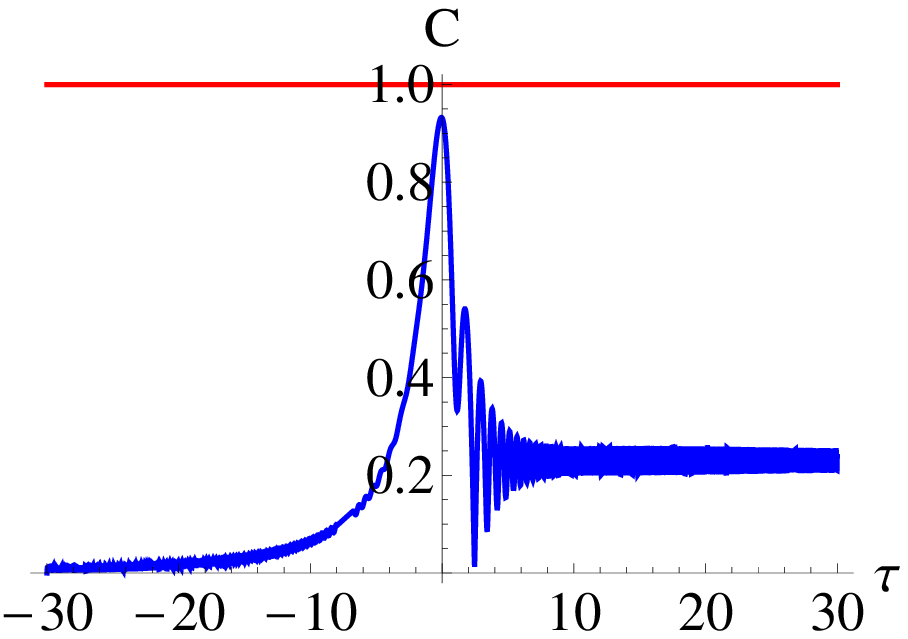}\label{fig:CB2beta05}}
\qquad
\subfloat[][]{\includegraphics[width=0.22\textwidth]{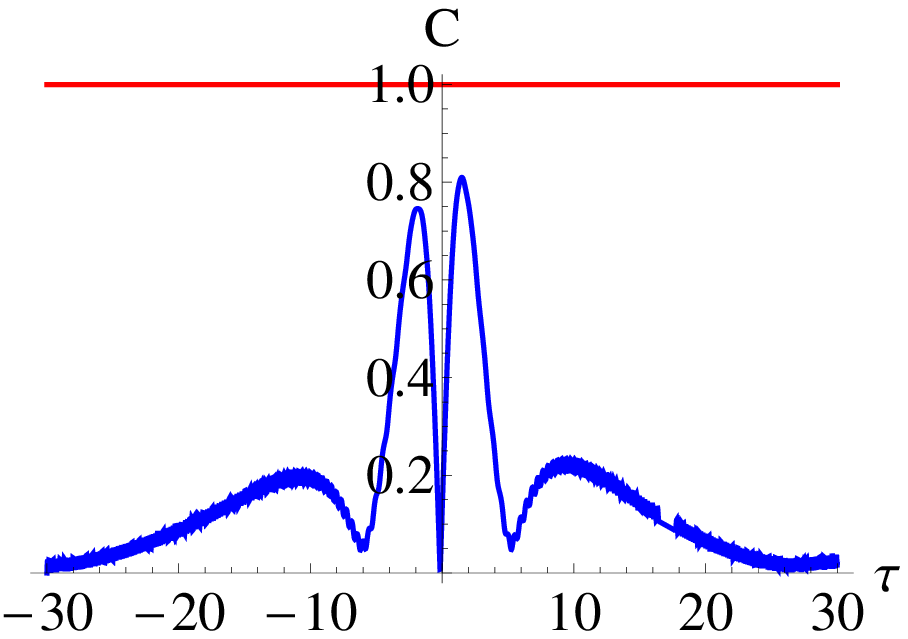}\label{fig:CB2beta2}}
\qquad
\subfloat[][]{\includegraphics[width=0.22\textwidth]{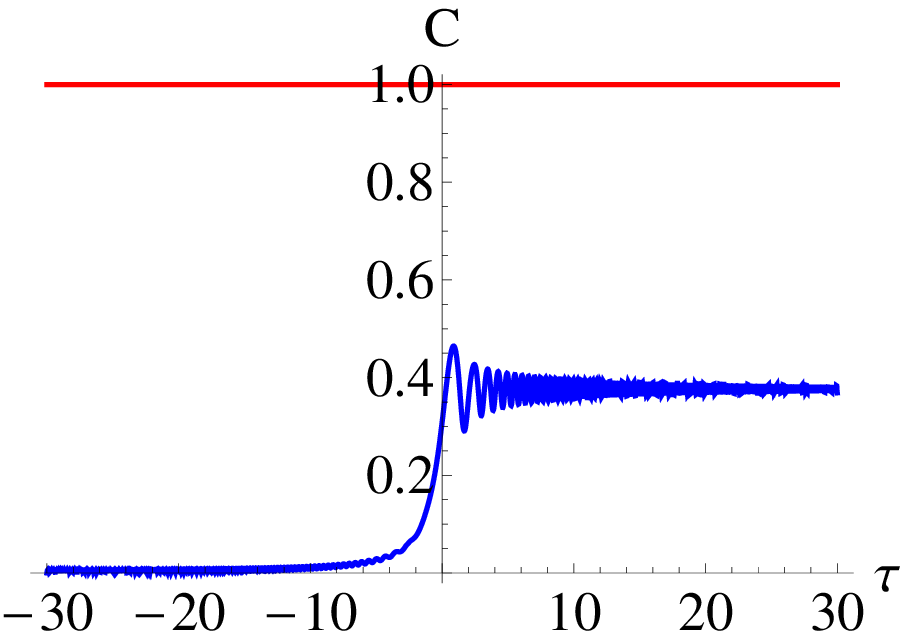}\label{fig:CB2beta01}}
\qquad
\subfloat[][]{\includegraphics[width=0.22\textwidth]{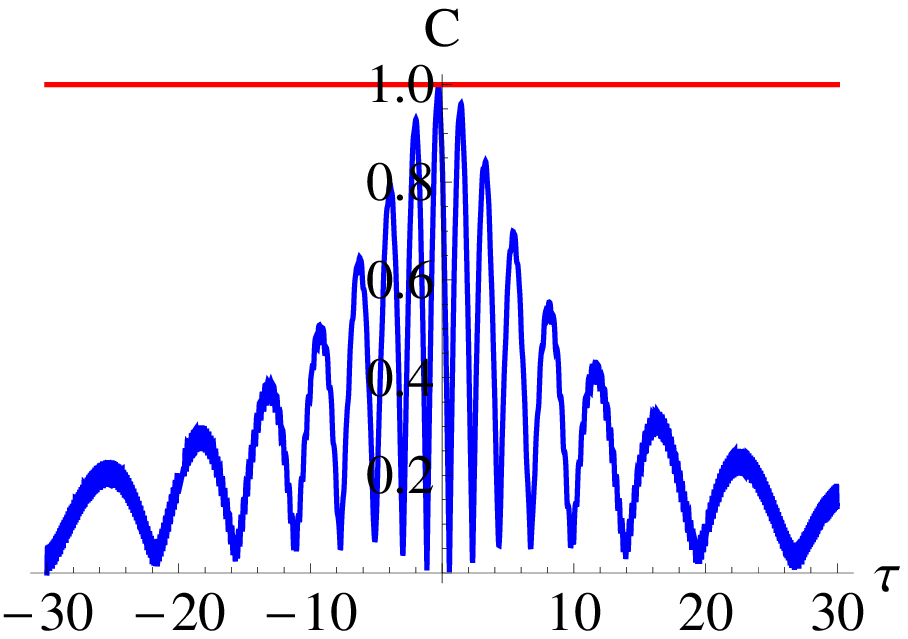}\label{fig:CB2beta10}}
\captionsetup{justification=raggedright,format=plain,skip=4pt}%
\caption{(Color online) Time behaviour of the concurrence against the dimensionless parameter $\tau=\sqrt{\alpha/\hbar}~t$ during a LMSZ process when the system starts from the state ($\ket{++}+\ket{+-})/\sqrt{2}$ for $\beta_-/\beta_+=1/2$ and a) $\beta_+=1/2$, b) $\beta_+=2$; $\beta_-/\beta_+=2$ and c) $\beta_+=0.5$, d) $\beta_+=2$;  $\beta_-/\beta_+=2$ and e) $\beta_+=0.1$, f) $\beta_+=10$. The upper straight curve corresponds to $C(\tau)=1$.}
\end{center}
\end{figure}

Such a ratio, conversely, results determinant for other initial conditions, e.g. the one considered in Eq. \eqref{Asymptotic In St 1 anisotropy}.
In this case the amplitudes read
\begin{equation}
c_{++}=a_+, \quad c_{--}=-b_+^*, \quad c_{+-}=a_-, \quad c_{-+}=-b_-^*.
\end{equation}
In Figs. \ref{fig:CB05beta05}-\ref{fig:CB2beta10} we may appreciate the influence of both the ratio $\beta_-/\beta_+$ and the free parameter $\beta_+$; the former influences only qualitatively the behaviour of the concurrence, while the latter both qualitatively and quantitatively.
This time too, the concurrence vanishes for high values of $\beta_+$ witnessing an asymptotic factorized state.
For small values of $\beta_+$, instead, positive values of entanglement even for large times indicate a superposition of the four standard basis states.
 
In conclusion of this section, we underline that in Ref. \cite{VitPRL2001} the authors considered a system of two spin-1/2's interacting only through the term $\hat{\sigma}_1^z\hat{\sigma}_2^z$ and subjected to the same magnetic field consisting in a Gaussian pulse uniformly rotating in the $x-y$ plane and a LMSZ ramp in the $z$ direction.
They showed that the coupling between the two spins enhances significantly the probability to drive adiabatically the two-spin system from the separate state $\ket{--}$ to the entangled state $(\ket{+-}+\ket{-+})/\sqrt{2}$.
In this case the procedure to generate an entangled state is different from the scenario considered here because of the different symmetries of the Hamiltonians ruling the two-spin dynamics.
Indeed, in Ref. \cite{VitPRL2001} the Hamiltonian commutes with $\hat{S}^2$ and consequently two dynamically invariant Hilbert subspaces exist: one of dimension three and the other of dimension one.
The three-dimensional subspace is spanned by the states $\ket{++}$, $(\ket{+-}+\ket{-+})\sqrt{2}$ and $\ket{--}$, making possible the preparation of the entangled state of the two spin-1/2's by an adiabatic passage when they start from the separate state $\ket{--}$.
In our case, instead, $\hat{S}^2$ is not constant while the integral of motion is $\hat{\sigma}_1^z\hat{\sigma}_2^z$.
The symmetries of the Hamiltonian, thus, generate two two-dimensional dynamically invariant Hilbert subspaces: one spanned by $\ket{++}$ and $\ket{--}$ and the other by $\ket{+-}$ and $\ket{-+}$.
Then, in our case, the transition between the states considered in the other work is impossible since such states belong to different invariant subspaces.

\section{Effects of Classical Noise}\label{Sec V}

In experimental physical contexts involving atoms, ions and molecules investigated and manipulated by application of lasers and fields, the presence of noise in the system stemming from the coupling with a surrounding environment is unavoidable.
Though a lot of technological progresses and experimental expedients have been developed, it is necessary to introduce such decoherence effects in the theoretical models for a better understanding and closer description of the experimental scenarios.
There exist different approaches to treat the influence of a thermal bath; one is to consider the presence of classical noisy fields \cite{Pok-Sin2003,Paris} stemming, e.g., from the presence and the influence of a surrounding nuclear spin bath \cite{Pok-Sin2003}.

In the last reference the authors study a noisy LMSZ scenario for a $N$-level system.
They take into account a time-dependent magnetic field $\eta(t)$ only in the $z$-direction and characterized by the following time correlation function $\average{\eta(t)\eta(t')}=2G\delta(t-t')$.
The authors show that the LMSZ transition probability $P_-^+$ for a spin-1/2 to be found in the state $\ket{+}$ starting from $\ket{-}$, in case of large values of $G$, changes as
\begin{equation}
P_-^+={1-\exp\{-2\pi g^2/\hbar\alpha\} \over 2},
\end{equation}
where $g$ is the energy contribution due to the coupling of the spin-1/2 with the constant transverse magnetic field and $\alpha$ is the ramp of the longitudinal magnetic field.
We see that the value of $G$, provided that it is large, does not influence the transition probability.
The unique effect of the noisy component is the loss of coherence.
The field indeed cannot generate transitions between the two diabatic states, being only in the same direction of the quantization axis.
In this way the transition probability, as reasonable, results hindered by the presence of the noise, since, for $g^2/\alpha\gg 1$, the system reaches at most the maximally mixed state. 

This result is of particular interest in our case since the addition of the noisy component $\eta(t)$ leaves completely unaffected the symmetry-based Hamiltonian transformation and the validity of the dynamics-decoupling procedure.
Thus, also in this case, the dynamical problem of the two-qubit system may be converted into two independent spin-1/2 problems affected by a random fluctuating $z$-field.
Thus, we may write easily the transition probabilities when the two spins are subjected to a unique homogeneous field influenced by the noisy component considered before.
We have precisely
\begin{equation}
P_+={1-\exp\{-2\pi \gamma_+^2/\hbar\alpha\} \over 2}, \quad \omega_1(t)=\omega_2(t)=[\alpha t+\eta(t)]/4,
\end{equation}
We underline that the transition probability $P_-$ vanishes in case of an unique homogeneous magnetic field.
In the related subdynamics, indeed, the effective field ruling the two-spin dynamics is zero, namely $\Omega_-(t)=0$.
Moreover, for $\gamma_x=\gamma_y$ we would have no physical effects, since, in such a case, also $P_+$ would result zero.

Another way to face with the problem of open quantum systems is to use non-Hermitian Hamiltonians effectively incorporating the information of the fact that the system they describe is interacting with a surrounding environment \cite{Feshbach,Moiseyev,Rotter,Simeonov,Torosov1,GdCKM}.
We may suppose, for example, that the spontaneous emission from the up-state to the down-one is negligible and that some mechanism makes the up-state $\ket{+}$ irreversibly decaying out of the system with rate $\xi$ and $\xi'$ for the first and second spin-1/2, respectively.
It is well known that we can phenomenologically describe such a scenario by introducing the non-Hermitian terms $i\xi\hat{\sigma}_1^z/2$ and $i\xi'\hat{\sigma}_2^z/2$ in our Hamiltonian model.
Analogously to the case of a noisy field component, also the introduction of these terms does not alter the symmetry of the Hamiltonian model.
The symmetry-based transformation leads us to two independent non-Hermitian two-level models.
In the same way we may exploit the results got for a single qubit with a decaying state subjected to the LMSZ scenario \cite{Akulin,Vitanov,Ivanov} and reread them in terms of the two-spin-1/2 language.
We know that the decaying rate affects only the time-history of the transition probability but not, surprisingly, its asymptotic value \cite{Akulin}.
However, this result is valid for the ideal LMSZ scenario; considering the more realistic case of a limited time window, it has been demonstrated, indeed, that a decaying rate-dependence for the population of the up-state arises \cite{Vitanov}.

\section{Conclusive Remarks}\label{Sec C}

In this work we considered a physical system of two interacting spin-1/2's whose coupling comprises the terms stemming from the anisotropic exchange interaction.
Moreover, each of them is subjected to a local field linearly varying over time.
The $C_2$-symmetry (with respect to the quantization axis $\hat{z}$) possessed by the Hamiltonian allowed us to identify two independent single spin-1/2 sub-problems nested in the quantum dynamics of the two spin-qubits.
This fact gave us the possibility of decomposing the dynamical problem of the two spin-1/2's into two independent problems of single spin-1/2.
In this way, our two-spin-qubit system may be regarded as a four-level system presenting an avoided crossing for each pair of instantaneous eigenenergies related to the two dynamically invariant subspaces.
This aspect turned out to be the key to solve easily and exactly the dynamical problem, bringing to light several physically relevant aspects.

In case of time-dependent Hamiltonian models, such a symmetry-based approach and the reduction to independent problems of single spin-1/2 has been used also in other cases \cite{GMN,GMIV,GBNM,GLSM}.
This fact permits a deep understanding of the quantum dynamics of the spin systems with consequent potential applications in quantum information and computation.
We underline, in addition, that the dynamical reduction exposed in Sec. \ref{Sec II} is independent of the time-dependence of the fields.
Thus, we may consider also different exactly solvable time-dependent scenarios \cite{KunaNaudts,Bagrov,Das Sarma,Mess-Nak,MGMN,GdCNM,SNGM} for the two subdynamics, resulting, of course, in different two-spin dynamics and physical effects.

In this paper, we showed that, although the absence of a transverse chirp \cite{VitPRL2001} or constant field, LMSZ transitions are still possible, precisely from $\ket{--}$ to $\ket{++}$ and from $\ket{-+}$ to $\ket{+-}$ (the two couples of states spanning the two dynamically invariant Hilbert spaces related to the symmetry Hamiltonian).
Such transitions occur thanks to the presence of the coupling between the spins which plays as effective static transverse field in each subdynamics.

It is worth noticing that, in our model, the two LMSZ sub-dynamics are ruled either by different combinations of the externally applied fields (when the local fields are different) or by the same field (under the STM scenario, that is when one local field is applied on just one spin).
In the latter case we showed the possibility of 1) a non-local control, that is to manipulate the dynamics of one spin by applying the field on the other one and 2) a state exchange/transfer between the two spins.
We brought to light how such effects are two different replies of the system depending on the isotropy properties of the exchange interaction.

Concerning the interaction terms, each subdynamics is characterized by different combinations of the coupling parameters.
This aspect has relevant physical consequences since, as showed, by studying the LMSZ transition probability in the two subspaces, it is possible both to evaluate the presence of different interaction terms and to estimate their weights in ruling the dynamics of the two-spin system.
We brought to light how the estimation of the coupling parameters could be of relevant interest since, through this knowledge, we may set the slope of variation of the LMSZ ramp as to generate asymptotically entangled states of the two spin-1/2's.
Moreover, we reported the exact time-behaviour of the entanglement for different initial conditions and we analysed how the coupling parameters can determine different entanglement regimes and asymptotic values.

Finally, we emphasized how our symmetry-based analysis has proved to be useful also to get exact results when a classical random field component or non-Hermitian terms are considered to take into account the presence of a surrounding environment interacting with the system.
In this case, the dynamics decomposition is unaffected by the presence of the noise or the dephasing terms and then we may apply the results previously reported for a two-level system \cite{Akulin,Vitanov,Pok-Sin2003} and reread them in terms of the two spin-1/2's.

We wish to underline, in addition, that our results are valid not only within the STM scenario, but they are applicable to other physical platforms.
Indeed, the local LMSZ model for a spin-qubit interacting with another neighbouring spin-qubit may be reproduced also in laser-driven cold atoms in optical lattices where highly-selective individual addressing has been experimentally demonstrated \cite{Weitenberg2011}.
Another prominent example is laser-driven ions in a Paul trap where spatial individual addressing of single ions in an ion chain has been routinely used for many years \cite{Monz2016,Bermudez2017}.
Yet another example is microwave-driven trapped ions in a magnetic-field gradient where individual addressing with extremely small cross-talk has been achieved in frequency space \cite{Piltz2014,Lekitsch2017}.

We point out that the results obtained in this paper are deeply different from the ones reported in other Refs. \cite{Reilly,Ribeiro,Ribeiro1} where systems of two spin-1/2's in a LMSZ framework have been investigated on the basis of an approximate treatment.
In these papers, indeed, the two spin-qubits are not directly coupled, but they interact through a common nuclear spin bath which they are coupled to.
Such a composite system behaves as a two-level system under several assumptions and to derive the effective single spin-1/2 Hamiltonian requires several approximations.
In Ref. \cite{Ribeiro1}, in particular, the effective Hamiltonian describes the coupling between the two-level system and a longitudinal time-dependent field which is not a pure LMSZ ramp, presenting a complicated functional dependence on the original Hamiltonian parameters.
There is, in addition, a time-dependent effective interaction between the two states possessing a complicated functional dependence on the confinement energy as well as the tunneling and Coulomb energies.
Although such an effective Hamiltonian goes beyond the standard LMSZ scenario, it may be considered similar to the LMSZ one since both Hamiltonians describe an adiabatic passage through an anticrossing.

In our case, instead, the two spin-1/2's are directly coupled, besides to be subjected to a random field stemming from the presence of a spin bath.
Furthermore, the effective two-state Hamiltonians governing the two-qubit dynamics in the two invariant subspaces are easily got without involving any assumption and/or approximation.
The two two-level Hamiltonians, indeed, are derived only on the basis of a transparent mathematical mapping between the two-qubit states in each subspace and the states of a fictitious spin-1/2.
Moreover, they describe exactly a LMSZ scenario with a standard avoided crossing where the transverse constant field is effectively reproduced by the coupling existing between the two qubits.
The treatment at the basis of this work remarkably enables us to explore peculiar dynamical aspects of the system described by Eq. \eqref{Hamiltonian}, leading, for example, to the exact evolution of the entanglement get established between the two spins.

We underline, moreover, that our study is not a special case of the one considered in Ref. \cite{Larson}, where a Lipkin-Meskow-Glick (LMG)  interaction model for $N$ spin-qubits subjected to a LMSZ ramp is considered.
The numerical results reported in Ref. \cite{Larson} are, indeed, based on the mean field approximation.
In addition, there is no possibility of considering in the LMG model effects stemming from the anisotropy between $x$ and $y$ interaction terms.

Finally, two challenging problems naturally extending the investigation here reported are 1) that considering the interaction of two qutrits \cite{GVM} in place of two qubits and 2) that taking into account the coupling of the two spins with a quantum baths \cite{SHGM} in place of the interaction with a classical random field.

\section*{Acknowledgements}
NVV acknowledges support  from the EU Horison-2020 project 820314 (MicroQC).
RG acknowledges economical supports by research funds difc 3100050001d08+, University of Palermo.


\begin{thebibliography}{99}

\bibitem{LMSZ}
L. D. Landau, Phys. Z. Sowjetunion \textbf{2}, 46 (1932);
E. Majorana, Nuovo Cimento \textbf{9}, 43 (1932);
E. C. G. St\"uckelberg, Helv. Phys. Acta \textbf{5}, 369 (1932);
C. Zener, Proc. R. Soc. London, Ser. A \textbf{137}, 696 (1932).

\bibitem{Rabi1937}
I. I. Rabi, Phys. Rev. \textbf{51} 652 (1937).

\bibitem{Vasilev}
G. S. Vasilev and S. S. Ivanov and N. V. Vitanov, Phys. Rev. A \textbf{75}, 013417 (2007).

\bibitem{Torosov}
B. T. Torosov and N. V. Vitanov, Phys. Rev. A \textbf{84}, 063411 (2011).

\bibitem{Vit-Garr}
N. V. Vitanov and B. M. Garraway, Phys. Rev. A \textbf{53}, 6 (1996).

\bibitem{AE}
L. Allen and J. H. Eberly, Optical Resonance and Two-Level Atoms (Dover, New York, 1987); F. T. Hioe, Phys. Rev. A \textbf{30}, 2100 (1984).

\bibitem{DK}
Y. N. Demkov and M. Kunike, Vestn. Leningr. Univ. Fiz. Khim. \textbf{16}, 39 (1969).

\bibitem{Vasilev1}
G. S. Vasilev and N. V. Vitanov, Phys. Rev. A \textbf{73}, 023416 (2006).

\bibitem{Letho}
J. M. S. Lehto and K.-A. Suominen, Phys. Rev. A \textbf{94}, 013404 (2016).

\bibitem{Shevchenko1}
S.N. Shevchenko, S. Ashhab, Franco Nori, Phys. Rep. \textbf{492}, 1 (2010).

\bibitem{SSIvanov}
S. S. Ivanov and N. V. Vitanov, Phys. Rev. A \textbf{77}, 023406 (2008).

\bibitem{Sinitsyn}
N. A. Sinitsyn, Phys. Rev. A \textbf{87}, 032701 (2013).

\bibitem{Militello1}
B. D. Militello and N. V. Vitanov, Phys. Rev. A \textbf{91}, 053402 (2015).

\bibitem{Thiele}
S. Thiele, F. Balestro, R. Ballou, S. Klyatskaya, M. Ruben, W. Wernsdorfer1, Science \textbf{344} 6188 (2014).

\bibitem{Reilly}
D. J. Reilly, J. M. Taylor, J. R. Petta, C. M. Marcus, M. P. Hanson, A. C. Gossard, Science \textbf{321}, 817 (2008).

\bibitem{Huang}
P. Huang, J. Zhou, F. Fang, X. Kong, X. Xu, C. Ju, and J. Du, Phys. Rev. X \textbf{1}, 011003 (2011).

\bibitem{Randall}
J. Randall, A. M. Lawrence, S. C. Webster, S. Weidt, N. V. Vitanov, and W. K. Hensinger1, Phys. Rev. A \textbf{98}, 043414 (2018).

\bibitem{Wieser}
R. Wieser, V. Caciuc, C. Lazo, H. H\"olscher, E. Y. Vedmedenko and R. Wiesendanger, New J. Phys. \textbf{15} 013011 (2013).

\bibitem{Sivkov}
I. N. Sivkov, D. I. Bazhanov and V. S. Stepanyuk, Sc. Rep. \textbf{7}, 2759 (2017).

\bibitem{Petta}
J. R. Petta  \textit{et al.}, Science \textbf{309} (5744) 2180-2184 (2005).

\bibitem{Anderlini}
M. Anderlini, P. J. Lee, B. L. Brown, J. Sebby-Strabley, W. D. Phillips, and J. V. Porto Nature \textbf{448} (7152) 452-456 (2007).

\bibitem{Foletti}
H. Bluhm, S. Foletti, I. Neder, M. Rudner, D. Mahalu, V. Umansky and A. Yacoby Nat. Phys. \textbf{7} 109 (2011).

\bibitem{Das Sarma Nat}
Xin Wang, L. S. Bishop, J. P. Kestner, E. Barnes, Kai Sun and S. Das Sarma Nat. Comm. \textbf{3} 997 (2012).

\bibitem{Akulin}
V. M. Akulin and W. P. Schleich, Phys. Rev. A \textbf{46}, 7 (1992).

\bibitem{Vitanov}
N. V. Vitanov and S. Stenholm, Phys. Rev. A \textbf{55}, 4 (1997).

\bibitem{Pok-Sin2003}
V. L. Pokrovsky and N. A. Sinitsyn, Phys Rev. B \textbf{67}, 144303 (2003).

\bibitem{Ivanov}
P. A. Ivanov and N. V. Vitanov, Phys. Rev. A \textbf{71}, 063407 (2005).

\bibitem{Militello2}
M. Scala, B. Militello, A. Messina and N. V. Vitanov, Phys. Rev. A \textbf{84}, 023416 (2011).

\bibitem{Militello3}
A. V. Dodonov, B. Militello, A. Napoli, and A. Messina, Phys. Rev. A \textbf{93}, 052505 (2016).

\bibitem{VitPRL2001}
R. G. Unanyan, N.V. Vitanov, and K. Bergmann, Phys. Rev. Lett. \textbf{87}, 137902 (2001).

\bibitem{Ribeiro}
H. Ribeiro and G. Burkard, Phys. Rev. Lett. \textbf{102}, 216802 (2009).

\bibitem{Ribeiro1}
H. Ribeiro, J. R. Petta, and G. Burkard, Phys. Rev. B \textbf{87}, 235318 (2013).

\bibitem{Rancic}
M. J. Ranci\'c, D. Stepanenko, Phys. Rev. B \textbf{94}, 241301(R) (2016).

\bibitem{Larson}
J. Larson, Eur. Phys. Lett. \textbf{107}, 30007 (2014).

\bibitem{Shevchenko}
S. N. Shevchenko, A. I. Ryzhov, and Franco Nori, Phys. Rev. B \textbf{98}, 195434 (2018).

\bibitem{Mhel}
S. Mhel, Phys. Rev. B \textbf{91}, 035430 (2015).

\bibitem{GMN}
R. Grimaudo, A. Messina, H. Nakazato, Phys. Rev. A \textbf{94}, 022108 (2016).

\bibitem{Khajetoorians}
A. A. Khajetoorians, J. Wiebe, B. Chilian, R. Wiesendanger, Science \textbf{332} (6033) 1062-1064 (2011).

\bibitem{Yan}
S. Yan, D.-J. Choi, J. A. J. Burgess, S. Rolf-Pissarczyk and S. Loth, Nature Nanotechnology \textbf{10}, 40–45 (2015).

\bibitem{Bryant}
B. Bryant, A. Spinelli, J. J. T. Wagenaar, M. Gerrits, and A. F. Otte, Phys. Rev. Lett. \textbf{111}, 127203 (2013).

\bibitem{Tao}
Kun Tao, V. S. Stepanyuk, W. Hergert, I. Rungger, S. Sanvito, and P. Bruno, Phys. Rev. Lett. \textbf{103}, 057202 (2009).

\bibitem{Wootters}
W. K. Wootters, Phys. Rev. Lett. \textbf{80}, 2245 (1998).

\bibitem{Abramowitz}
M. Abramowitz and I. A. Stegun, \textit{Handbook of Mathematical Functions} (Dover, New York, 1964).

\bibitem{Paris}
C. Benedetti and M . Paris, Phys. Lett. A \textbf{378} 2495–2500 (2014).

\bibitem{Feshbach}
H. Feshbach, Ann. Phys. \textbf{5}, 357 (1958); H. Feshbach, Ann. Phys. \textbf{19}, 287 (1962).

\bibitem{Moiseyev}
N. Moiseyev, Non-Hermitian Quantum Mechanics (Cambridge Univ. Press, Cambridge, 2011).

\bibitem{Rotter}
I. Rotter and J. P. Bird, Rep. Prog. Phys. \textbf{78}, 114001, (2015);
I. Rotter, J. Phys. A: Math. Theor. \textbf{42}, 153001, (2009).

\bibitem{Simeonov}
L. S. Simeonov and N. V. Vitanov, Phys. Rev. A \textbf{93}, 012123 (2016).

\bibitem{Torosov1}
B. T. Torosov and N. V. Vitanov, Phys Rev. A \textbf{96}, 013845 (2017).

\bibitem{GdCKM}
R. Grimaudo, A. S. M. de Castro, M. Ku\'s, and A. Messina, Phys. Rev A \textbf{98}, 033835 (2018).

\bibitem{GMIV}
R. Grimaudo, A. Messina, P. A. Ivanov and N. V. Vitanov J. Phys. A \textbf{50} 175301 (2017).

\bibitem{GBNM}
R. Grimaudo,Y. Belousov, H. Nakazato and A. Messina, Ann. Phys. (NY) \textbf{392}, 242 (2017).
    
\bibitem{GLSM}
R. Grimaudo, L. Lamata, E. Solano, A. Messina, Phys. Rev. A \textbf{98}, 042330 (2018).

\bibitem{Bagrov}
V. G. Bagrov, D. M. Gitman, M. C. Baldiotti and A. D. Levin, Ann. Phys. (Berlin) \textbf{14} (11) 764 (2005).

\bibitem{KunaNaudts}
M. Kuna and J. Naudts Rep. Math. Phys. \textbf{65} (1) 77 (2010).

\bibitem{Das Sarma}
E. Barnes and S. Das Sarma Phys. Rev. Lett. \textbf{109} 060401 (2012).

\bibitem{Mess-Nak}
A. Messina and H. Nakazato J. Phys. A: Math. Theor. \textbf{47} 445302 (2014).

\bibitem{MGMN}
L. A. Markovich, R. Grimaudo, A. Messina and H. Nakazato Ann. Phys. (NY) \textbf{385} 522 (2017).

\bibitem{GdCNM}
R. Grimaudo, A. S. M. de Castro, H. Nakazato and A. Messina, Ann. Phys. (Berlin) \textbf{530}, 12 1800198 (2018).

\bibitem{SNGM}
T. Suzuki, H. Nakazato, Roberto Grimaudo and Antonino Messina, Sc. Rep. \textbf{8}, 17433 (2018).

\bibitem{Weitenberg2011}
C. Weitenberg, M. Endres, J. F. Sherson, M. Cheneau, P. Schauss, T. Fukuhara, I. Bloch, and S. Kuhr,
Nature \textbf{471}, 319 (2011).

\bibitem{Monz2016}
T. Monz, D. Nigg, E. A. Martinez, M. F. Brandl, P. Schindler, R. Rines, S. X. Wang, I. L. Chuang, R. Blatt,
Science \textbf{351}, 1068 (2016).

\bibitem{Bermudez2017}
A. Bermudez, X. Xu, R. Nigmatullin, J. O’Gorman, V. Negnevitsky, P. Schindler, T. Monz, U. G. Poschinger, C. Hempel, J. Home, F. Schmidt-Kaler, M. Biercuk, R. Blatt, S. Benjamin, and M. M\"uller,
Phys. Rev. X \textbf{7}, 041061 (2017).

\bibitem{Piltz2014}
C. Piltz, T. Sriarunothai, A. F. Varon, and C. Wunderlich,
Nature Commun. \textbf{5}, 4679, (2014).

\bibitem{Lekitsch2017}
B. Lekitsch, S. Weidt, A. G. Fowler, K. M{\o}lmer, S. J. Devitt, C. Wunderlich, and W. K. Hensinger,
Science Advances \textbf{3}, e1601540 (2017).

\bibitem{GVM}
R. Grimaudo, N. V. Vitanov, A. Messina, arXiv:1901.00322v1 (2019).

\bibitem{SHGM}
A. Sergi, G. Hanna, R. Grimaudo and A. Messina, Symmetry, \textbf{10}(10), 518 (2018).

\end{thebibliography}
\end{document}